# Simulation of the Exophthalmia Reduction using a Finite Element Model of the Orbital Soft Tissues

Vincent Luboz[1], Annaig Pedrono[2], Pascal Swider[2], Frank Boutault[3] and Yohan Payan[1]

[1] Laboratoire TIMC-GMCAO, UMR CNRS 5525, Faculté de Médecine Domaine de la Merci, 38706 La Tronche, France
{Vincent.Luboz, Yohan.Payan}@imag.fr
[2] INSERM, Biomécanique, CHU Purpan, BP 3103, 31026 Toulouse Cedex 3, France
{apedrono, pascal.swider}@toulouse.inserm.fr
[3] CHU Purpan, Service de Chirurgie Maxillo-faciale, 31026 Toulouse Cedex 3, France
BOUTAULT.F@chu-toulouse.fr

**Abstract.** This paper proposes a computer-assisted system for the surgical treatment of exophthalmia. This treatment is classically characterized by a decompression of the orbit, by the mean of an orbital walls osteotomy. The planning of this osteotomy consists in defining the size and the location of the decompression hole. A biomechanical model of the orbital soft tissues and its interactions with the walls are provided here, in order to help surgeons in the definition of the osteotomy planning. The model is defined by a generic Finite Element poro-elastic mesh of the orbit. This generic model is automatically adapted to the morphologies of four patients, extracted from TDM exams. Four different FE models are then generated and used to simulate osteotomies in the maxillary or ethmoid sinuses regions. Heterogeneous results are observed, with different backwards movements of the ocular globe according to the size and/or the location of the hole.

## 1 Introduction

The exophthalmia is a pathology characterized by an excessive forward displacement of the ocular globe outside the orbit (Figure 1 (a)). This forward displacement ("protrusion") is a consequence of an increase in the orbital content behind the globe. The consequences of the exophthalmia are aesthetical and psychological. It may have functional consequences such as a too long cornea exposition or, in the worst case, a distension of the ocular nerve that leads to a decrease of the visual acuity and sometimes to a total blindness. Four origins can be found to exophthalmia [1]. First, following a trauma, an haematoma can compressed the optic nerve which can reduce the visual perception. A surgical decompression of the haematoma may then be necessary. The second exophthalmia cause is cancerous, with tumor in one or both orbits that may reduce the mobility of the globe. Radiotherapy can be used and surgical extraction of the tumor is sometimes needed. Third cause of exophthalmia: infections, that are treated with antibiotics. Finally, and for most cases, exophthalmia can be due to endocrinal dysfunction, such as the Basedow illness which is related to a thyroid pathology. This cause often leads to a bilateral exophthalmia as this dysfunction induces an increase of the ocular muscles and fat tissues volume. Once the endocrinal situation is stabilized, a surgical reduction of the exophthalmia is usually needed.

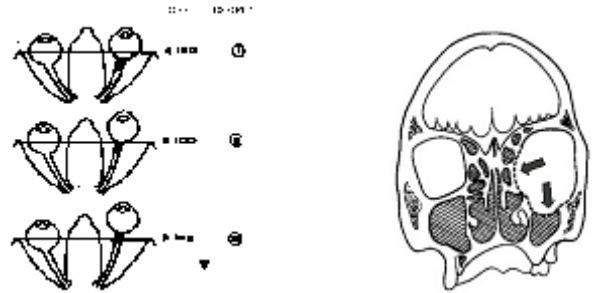

**Fig. 1.** (a) Left: three degrees of exophthalmia (from top to bottom, the protrusion becomes more and more severe). (b) Right: decompression in the sinuses region; some fat tissues are evacuated inside those new cavities.

Two surgical techniques are classically used for this decompression. The first one aims at extracting few cm$^3$ of fat tissues via an external eyelid incision. The advantage of this method is the relative security during surgery since the ocular nerve is far from the fat tissues that are extracted. The drawbacks are (1) a small backward displacement of the globe due to the limited fat tissues in the eyelid region, and (2) an aesthetical risk for the patient since the eyelid incisions can leave visible scars. The second surgical technique aims at increasing the volume of the orbital cavity, with an osteotomy of the orbital walls [2][3]. No fat tissues are therefore extracted. Indeed, following this osteotomy (*i.e.* a hole in the maxillary or ethmoid sinuses regions Figure 1 (b)) and with an external pressure exerted by the surgeon onto the ocular globe, some fat tissues can be evacuated through the hole, towards the sinuses. The advantages of this method are (1) limited scars due to the incision that is made inside the eyelid region and (2) a backward displacement of the ocular globe that can be much higher. The drawback is the risk for the surgeon to cause a decrease of the visual acuity or even a blindness since he works near the ocular nerve. The surgical gesture has therefore to be very precise to avoid critical structures. Moreover this intervention is technically difficult since the eyelid incision is narrow and gives to the surgeon a small visibility of the operating field.

This gesture could be assisted in a computer-aided framework from two points of view. First, a computer-guided system could help the surgeon in localizing the tip of the surgical tool inside the orbit and therefore maximize the precision of the intervention. Second, a model could assist the surgeon in the planning definition: where opening the orbit walls and to which extend? These two questions are directly related to the surgical objective that is expressed in term of suited backward displacement of the ocular globe. This paper addresses the second point, by proposing a biomechanical 3D Finite Element (FE) model of the orbit, including the fat tissues and the walls. First qualitative simulations of walls opening are provided on different patient models and compared with observed phenomena.

## 2. Orbital soft tissue modeling and surgical simulation

### 2.1 Strategy

The modeling strategy that has been adopted by our group is driven by the surgeons needs: (1) the models must be precise as they are used for clinical predictions and (2) tools must be provided to automatically adapt models to each patient morphology. Our strategy to follows those requirements is (see [4] for a review):
- the manual elaboration of a "generic" ("atlas") biomechanical model; this step is long, tedious, but is a prerequisite for accuracy of the modeling;
- patient data acquisition, from US, CT or MRI exams;
- conformation of the generic biomechanical model to patient morphology, by the mean of elastic registration.

In the case of exophthalmia, pre-operative and post-operative CT exams are systematically collected. This modality is therefore used for the elaboration of the generic model, as well as for its conformation towards patient geometries.

### 2.2 A generic biomechanical model of the orbit cavity

In order to define the generic model of the orbital soft tissues, CT data, collected for a given patient (considered therefore as the "generic" patient), are used for the extraction of the orbital cavity, fat tissues, ocular muscles and nerve geometries.

#### 2.2.1 Tri-dimensional reconstruction of the orbit cavity
For each image of the CT exam, a manual segmentation of the main ocular structures is made through B-Splines definitions (figure 2 (a)). By connecting those curves, 3D reconstruction is provided for the orbital cavity, the muscles as well as the ocular nerve (figure 2 (b)).

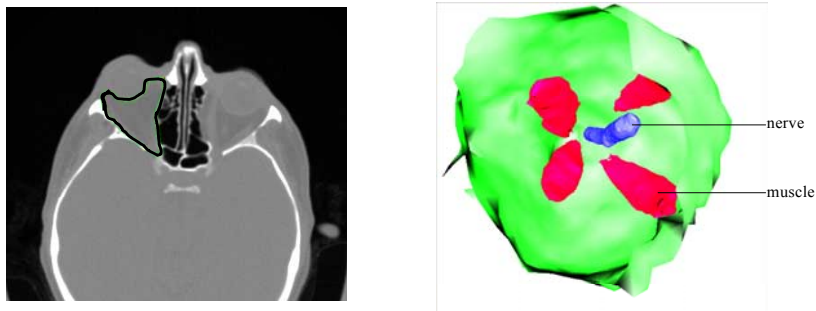

**Fig. 2.** (a) Manual Spline segmentation. (b) 3D representation of the orbital cavity, with the orbit limits, the four muscles and the ocular nerve (the ocular globe is not displayed).

In addition to the 3D reconstruction, this image processing tool can be used to compute volumes of individual ocular structures. For example, in the case of the patient used to define the geometry of the "generic" model, a value of 30cm$^3$ is measured for the orbital cavity.

### 2.2.2 Biomechanical modeling of the orbital structures

Besides the study of the intra orbital soft tissues volumes, the segmentation of the CT exam provides a base for the definition of a 3D Finite Element mesh. To our knowledge, the human orbit has not yet been modeled through a FE approach. The ocular globe has been studied using a FE model [5] but this model is not appropriate for the exophthalmia study since it does not include the soft tissues behind the globe.

Starting from the successive Splines segmented from CT slices, a volumetric mesh, made of hexahedrons, has been manually built. Because of the huge amount of work needed for this meshing, the internal soft tissues, namely the fat, muscles and nerve, were all modeled as a single biomechanical entity.

The biomechanics of this entity was modeled with the FE Method considering a poro-elastic behavior of the structure. This material was chosen following discussions with clinicians that describe the intra-orbital soft tissues as similar to the behavior of a sponge full of water (a kind of mixture between unorganized fibers and fluids).

The MARC™ Finite Element package was used to model this material, with mechanical parameters values closed to soft tissues values reported in the literature [6]: 20kPa for the Young modulus, 0.4 for the Poisson's ratio. A 0.01mm$^4$/N.s value was finally chosen for the permeability, in order to model the observed high retention of the fluid.

### 2.2.3 Simulation of the surgical gesture

In order to simulate the surgical gesture characterized by (1) an osteotomy of the orbital walls and (2) a pressure exerted by the surgeon onto the ocular globe, specific boundary conditions were defined for the generic model.

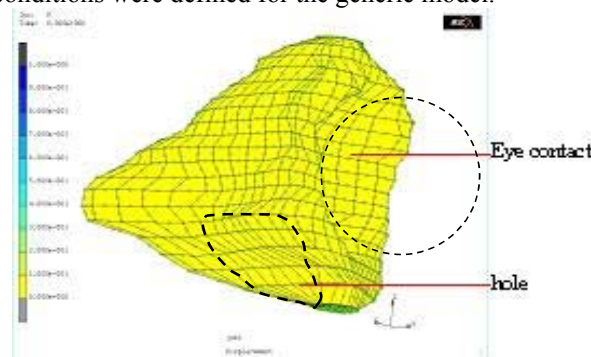

**Fig. 3.** 3D FE generic mesh: all elements are hexahedrons ; the simulated hole is located in the ethmoid sinus region.

First, to simulate the osteotomy, nodes located in the hole region (figure 3) were assumed to be free to move, whereas the other surface nodes located along the orbital walls were maintained fixed. All the other nodes located inside the volumetric mesh were modeled without any boundary conditions. Second, in order to model contact forces between internal soft tissues and the ocular globe, resulting from the pressure exerted by the surgeon, an imposed force of 10 Newton was distributed along the nodes located at the interface (figure 3).

**2.3 Conformation of the generic model to patient morphology**

Following the strategy introduced in part 2.1, data has to be collected for each patient. CT exam being classically required for exophthalmia, surfacic points, located along the orbital walls, were manually extracted from CT images, by the mean of the manual B-Spline segmentation process described above. A set of 3D points, describing the patient orbit surface geometry, is thus obtained. The Mesh-Matching algorithm [4], followed by a correction method [7] is then applied to match the generic model towards the patient geometry. A new 3D mesh is therefore automatically generated, adapted to patient geometry and sufficiently regular to perform a FE Analysis.

## 3. Results

**3.1 Patient conformation and simulation of the orbital decompression**

The aim of this study is to assist the surgical planning by estimating (1) the influence of the hole size and of its location and (2) the mechanical behavior of the orbit, with the simulated soft tissues evacuation through the hole. Four patients were studied for this paper. In collaboration with surgeons, and for each patient, four different holes were simulated, assuming two locations (the forwards and backwards ethmoid sinus regions) and two sizes for the degree of osteotomy (1.4cm² and 2.9cm²).

Figure 4 plots one patient FE model generated by the conformation process, and the corresponding four holes. For these figures, a 10 Newton pressure force values was applied to the globe.

Two aspects of the simulations were studied: (1) the relation between the hole size/location and the ocular globe backward displacement and (2) the observed decompression of the orbit, with the mechanical behavior of the soft tissues.

**3.2 Relation between hole size/location and ocular globe backward displacement**

For each of the four patients, simulations were carried out and relationships between the exerted globe pressure and the corresponding simulated backward displacement were computed.

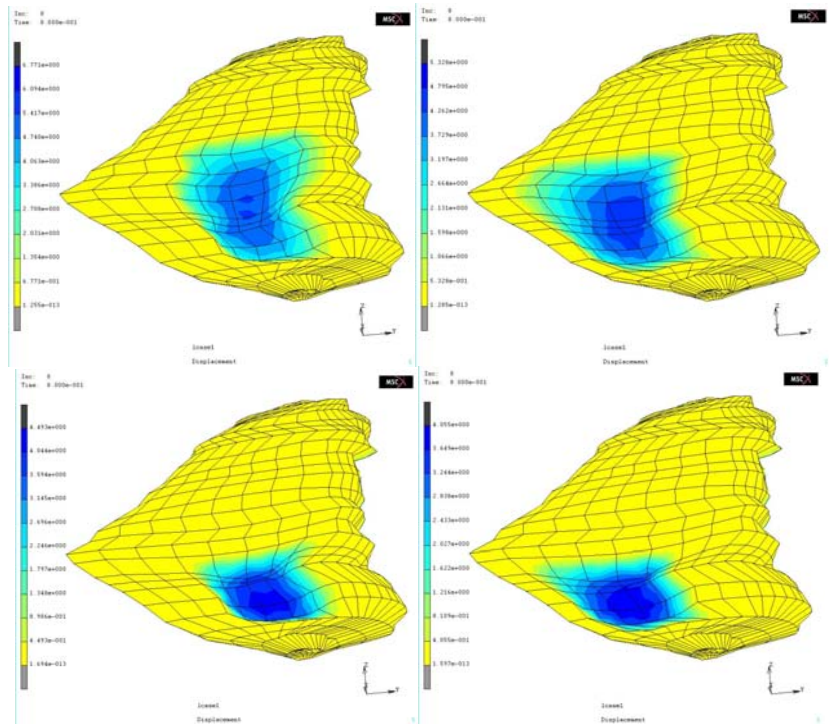

**Fig. 4.** Patient simulations with four different holes, from left to right: large holes at the front of the orbit and at the back of the orbit (top), and medium holes at the front of the orbit and at the back of the orbit (bottom).

The first interesting point is that, whatever the patient, all relationships are very similar. Figure 5 plots the relationship for one given patient. Whatever the level of exerted pressure force, the globe displacements due to the big hole located at the front of the orbit are 10% greater than displacements observed with a backward hole, and 40% greater than displacements due to the medium holes. Two consequences can be addressed by those results. First, both medium holes seem not able to provide a globe backward displacement larger than 3mm. Second, the big hole located at the front of the orbit seems more "efficient" as it provides a greater backward displacement of the globe. For example, with a 0.02MPa pressure value (which correspond to a force of 6N applied onto the globe), a backward displacement of nearly 5mm is reached with this hole. Those results seem to be realistic as they stand in the range of values reported by [8]. Of course, clinical measurements are needed to confirmed these analysis. Note also that the time required for a FE simulation with our model still does not fulfill the needs of the surgeon. Indeed, on a PC equipped with a 1,7GHz processor and a 1Go memory, the process lasts nearly 2 hours which is far from the real-time model needed for a computer assisted surgery.

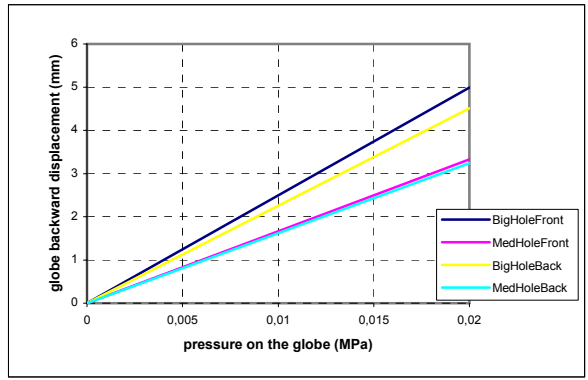

**Fig. 5.** Globe backward displacement resulting from various pressure forces (simulations are carried out for a given patient and with the four holes describe above).

### 3.3 Decompression of the orbit

This part evaluates the mechanical behavior of the intrinsic soft tissues, with in particular the total amount of tissues that are evacuated through the hole. Figure 6 plots, for each patient, the computed evacuated volume under different levels of pressure forces. The interesting observation is that, whereas few heterogeneity among patients was observed for the globe backward displacement, significant variations of the volume of fat tissues evacuated through the hole are obtained from one patient to the other one. Those variations are probably due to the shape of the FE mesh and therefore to the geometry of each patient orbit. For example, a difference of 30% can be seen between patientZJ and patientFJ, for the same hole and the same pressure.

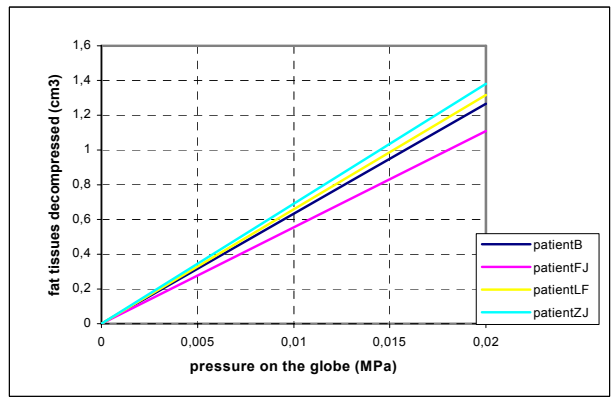

**Fig. 6.** Volume of fat tissues decompressed and evacuated through the large hole located at the front of the orbit. The volume is computed for the four patients.

## 3 Conclusion

This paper has presented a first evaluation of the feasibility of a Finite Element modeling of the surgical orbital decompression, in the context of exophthalmia. First simulations were carried out on four patients. Finite Element models of these patients were automatically generated from an existing generic model. For each patient, four different holes were simulated in the ethmoid sinus regions: two locations and two sizes. The ocular globe backward displacements as well as the intrinsic soft tissues that are evacuated through the hole were carefully studied, as they can be both considered as criteria for the "efficiency" of the surgery. Heterogeneity is the main conclusion of these simulations. Indeed, different levels of globe backward displacement are observed according to the sizes and/or to the locations of the holes. Moreover, for a given hole size and location, different volumes for soft tissues evacuated through this hole are observed according to patients geometries.

It seems therefore that surgeons should use such results in order to optimize their surgical planning. Before such a perspective, two points must be addressed.

First, the method must be clinically validated, by comparing simulations provided by the model, with pre and post-operative data collected on a given patient. Second, computation times must be studied and drastically improved, in order to provide a solution that allows fast simulations.